\documentclass[aps,prl,twocolumn]{revtex4}
\usepackage{graphicx}
\usepackage{amsmath}
\usepackage{amssymb}
\usepackage{epsfig}
\newcommand{\ket}[1]{\lvert#1\rangle}

\begin{document}

\title{Transition from Free to Interacting Composite Fermions away from $\nu=1/3$.}

\author{Y. \surname{Gallais}$^{1}$}
\email{yann@phys.columbia.edu}
\author{T.H. \surname{Kirschenmann}$^{1}$}
\author{I. \surname{Dujovne}$^{1}$}
\altaffiliation{Delft University of Technology, Kavli Insitute of NanoScience, Delft, The Netherlands}
\author{C.F. \surname{Hirjibehedin}$^{1}$}
\altaffiliation{IBM Ressearch Division, Almaden Research Center, San Jose, CA 95120}
\author{A. \surname{Pinczuk}$^{1,2}$}
\author{B.S. \surname{Dennis}$^{2}$}
\author{L.N. \surname{Pfeiffer}$^{2}$}
\author{K.W. \surname{West}$^{2}$}
\affiliation{$^1$Departments of Physics and of Applied Physics, Columbia University, New York, NY 10027 \\
$^2$Bell Labs, Lucent Technologies, Murray Hill, NJ 07974}

\begin{abstract}
Spin excitations from a partially populated composite fermion level
are studied above and below $\nu=1/3$. In the range $2/7<\nu<2/5$
the experiments uncover significant departures from the
non-interacting composite fermion picture that demonstrate the
increasing impact of interactions as quasiparticle Landau levels are
filled. The observed onset of a transition from free to interacting
composite fermions could be linked to condensation into the higher
order states suggested by transport experiments and numerical
evaluations performed in the same filling factor range.
\end{abstract}

\maketitle

In the composite fermion (CF) picture of the fractional quantum
Hall effect (FQHE) fundamental interactions are taken into
account at the mean field level by mapping the system of strongly
interacting electrons in a perpendicular magnetic field into a a
system of composite fermions (CF) moving in a reduced effective
magnetic field \cite{Jain}. The effective magnetic field
experienced by CF quasiparticles is $B^*$=$B-B_{\frac{1}{\phi}}$,
where $\phi$ is an even integer that labels different sequences of
incompressible states and $B_{\frac{1}{\phi}}$ is the magnetic
field corresponding to the filling factor $\nu$=$\frac{1}{\phi}$.
In the CF framework the reduction in magnetic field follows from the binding of $\phi$
flux quanta to electrons. The main sequence of FQHE states
are at Landau level filling factor $\nu$=p/($\phi$p$\pm$1), where
$p$ is the CF filling factor. Each FQHE sequence is then centered
around an even-denominator fraction, $\nu=1/\phi$, where the
effective field cancels and a compressible Fermi liquid is thought
to form \cite{HLR}. Underlying the CF paradigm is the formation of
CF Landau levels with spacing that depends on the effective field
$B^*$ as
\begin{equation}
\omega_{CF}=\frac{e\left|{B^*}\right|}{m^{*}c}
\label{CF}
\end{equation}
where $m^{*}$ is an effective CF mass.
\par
\begin{figure}[b]
\centering \epsfig{figure=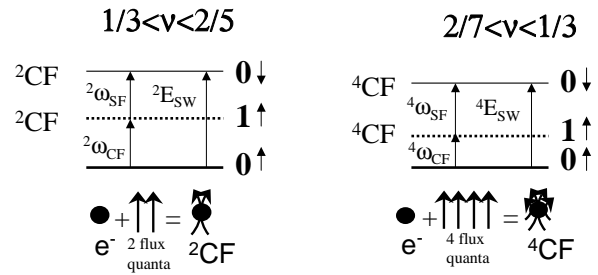, width=0.89\linewidth,clip=}
\caption{Composite Fermion level structure for $\nu>1/3$ ($^2$CF or
$\phi$=2 flux attachment, left) and  $\nu<1/3$ ($^4$CF or $\phi$=4
flux attachement, right) with the corresponding spin excitations:
the spin-flip (SF) and the spin-wave (SW). The dotted lines
represent the partially populated levels.} \label{fig1}
\end{figure}

In the CF framework the complex strongly interacting many-body
physics is transformed into the simple picture of non-interacting CF
at filling factors $p$. The sequence of lowest spin-split CF Landau
levels is shown in Fig. \ref{fig1} for quasiparticles with $\phi$=2
(or $^2$CFs) and $\phi$=4 (or $^4$CFs). The issue of the impact of
residual CF interactions is of particular relevance to states that
have $1<p<2$ in which there is partial population $p^*=p-1$ of an
excited Landau level (see Fig. \ref{fig1}). Residual CF interactions
in the partially populated excited CF level are believed to be
responsible for the formation of higher order FQHE states such as
$\nu=4/11$ and $\nu=4/13$ that have $p^*=1/3$ for $^2$CFs and
$^4$CFs \cite{Pan}. The experimental evidence for such higher order
states bring into question the validity of the weakly interacting CF
framework \cite{Quinn}. While these higher order fractions were
actually predicted in the earlier hierarchical model
\cite{Haldane,Halperin}, it was argued that the CF model may account
for these new fractions when CF residual interactions are
incorporated beyond the mean-field level
\cite{Wojs-Quinn,Jain-Mandal,Mandal-Jain,Goerbig1,Goerbig2,Jain-chang,
Lopez-Fradkin,Wojs-Quinn2}.
\par
Experimental evidence for the formation of a Landau level structure
of CFs is found in light scattering measurements of spin-flip (SF)
excitations in which Landau level number and orientation of spin
change simultaneously \cite{Irene1}. At filling factors away from
$\nu=1/3$, the lowest lying SF excitations probe the level structure
because the transitions in these excitations emanate from the partially
populated CF level as shown in Fig. \ref{fig1} \cite{Murthy,Mandal}.
Since SF excitations can be observed at fractional values
of $p$, experiments that probe lowest lying SF modes are
powerful tools in studies of CF interactions at filling factors away
from the major FQHE states.
\par
In this Letter we report measurements of low lying SF excitations
by inelastic light scattering that reveal marked quasiparticle
interactions when there is partial population of the $|1,\uparrow>$
CF Landau level ($0<p^*<1$). The experiments indicate a sharp
onset of CF interactions that occurs for
$p^*\approxeq1/5$. These results can be regarded as evidence that
emergence of higher order FQHE states away from filling factor
$\nu=1/3$ is linked to a transition from 'free' to interacting
behavior at relatively low occupation of the $\ket{1,\uparrow}$ CF
level.
\par
The sensitivity of low lying SF modes to interactions among CF
quasiparticles occurs because in a 'single-particle', or 'free',
CF paradigm (no impact of CF interactions), the mode linewidths
are expected to be independent of $p^*$ and their light scattering
intensities should be proportional to $p^*$. We find that at low
values of $p^*<0.2=1/5$, the SF mode remains sharp (width
$\gamma\leq0.06~meV$) and its light scattering intensity indeed is
linear in $p^*$ for both $\nu>1/3$ and $\nu<1/3$. When $p^*>1/5$,
however, the light scattering spectra reveal marked departures
from the 'free' CF behavior. The SF mode of $^2$CFs displays
significant broadening, with $\gamma$ reaching 0.1~meV for
$p^*>1/2$, and the light scattering intensities of SF modes of
$^2$CFs and $^4$CFs saturate when $p^* \rightarrow 1/2$.
\par
The broadening of the light scattering peak of SF modes for
$p^*>1/5$ suggests a small energy scale for $^2$CFs interactions
that is in the range of 0.1~meV. It is conceivable that
the broadening is linked to lower lying collective excitations
that exist when there is partial population of the
$\ket{1,\uparrow}$ $^2$CF level. In this scenario the energy range
for such excitations would be comparable to the enhanced
broadening of SF excitations. Currently there is no evaluation of
SF excitations as a function of $p^*$ but the observed broadening energies
are in the range of predictions for the energy scale of residual
CF interactions \cite{Goerbig1,Jain-chang}.

\par
\begin{figure}
\centering \epsfig{figure=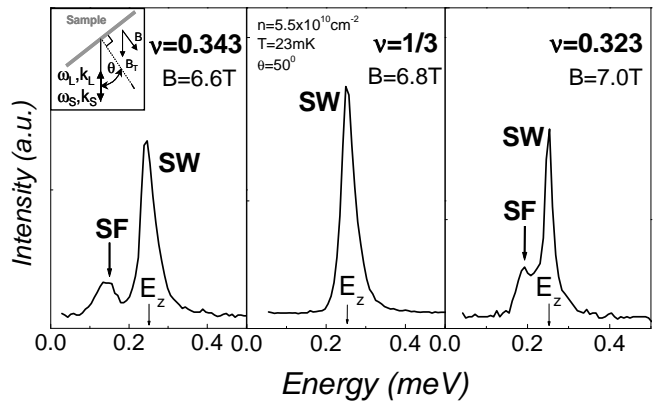, width=0.99\linewidth,clip=}
\caption{Light scattering spectra of the low-lying spin
excitations at three different filling factors: $\nu=0.343$, $\nu=1/3$
 and $\nu=0.323$. The scattering
geometry is shown in inset.} \label{fig2}
\end{figure}

The inelastic light scattering measurements were performed on a
high quality GaAs single quantum well (SQW) of width 330~$\AA$.
The electron density is n=5.5x10$^{10}$~cm$^{-2}$ and its low
temperature mobility is $\mu$=7.2x10$^6$~cm$^2$/V s.The sample was
mounted on the cold finger of a $^3$He/$^4$He dilution
refrigerator with windows for optical access that is inserted in
the cold bore of a superconducting magnet. Cold finger
temperatures can reach as low as T=23~mK. The back scattering
geometry was used at an angle $\theta$ with the normal of the
sample surface as shown in the inset to Fig. \ref{fig2}. The
magnetic field perpendicular to the sample is B=B$_T$cos$\theta$
and B$_T$ is the total field. The results reported here have been
obtained with $\theta$=50$^\circ\pm$2. Resonant inelastic light
scattering spectra were obtained by tuning the incident photon
energy of a Ti-Sapphire laser close to the fundamental optical gap
of GaAs to enhance the light scattering cross-section.The power
density was kept below 10$^{-5}$~W/cm$^2$ to prevent heating of
the electron gas. The scattered signal was dispersed by a triple
grating spectrometer working in additive mode and analyzed by a
CCD camera with 26$\mu m$ wide pixels. With a slit width of
30~$\mu m$, the combined resolution of the system is about
0.02~meV.
\par
Figure \ref{fig2} shows that two low-lying spin excitations are
observed. In addition to the SF mentioned above, there is the
spin-wave excitation (SW) in which only spin orientation changes
($\delta$S$_z$=$\pm$1)(see Fig. \ref{fig1}). By virtue of the Larmor
theorem, at long wavelengths the spin-wave is at the 'bare' Zeeman
energy E$_z$=g$\mu_BB_T$, where $\mu_B$ is the Bohr magneton and g
is the Land\'{e} factor of GaAs. On the other hand the energies and
lineshapes of SF excitations incorporate quasiparticle interactions
that offer probes of composite fermion physics.
\cite{Murthy,Mandal,Irene1,Irene2,Cyrus-thesis,Gallais}. At $\nu=1/3$, the only
low-energy spin excitation observed is the long wavelength SW mode
at E$_z$. When the electron fluid becomes compressible, for
\textit{both} $\nu>1/3$ and $\nu<1/3$, the SF peak appears on the
lower energy side of the SW mode. The narrow linewidth of SF
excitations on both sides of $\nu=1/3$ (Full Width at Half Maximum
or $\gamma<0.06~meV$) suggests that $\ket{1,\uparrow}$ Landau levels
of $^2$CFs and $^4$CFs are sharp.
\par
\begin{figure}
\centering \epsfig{figure=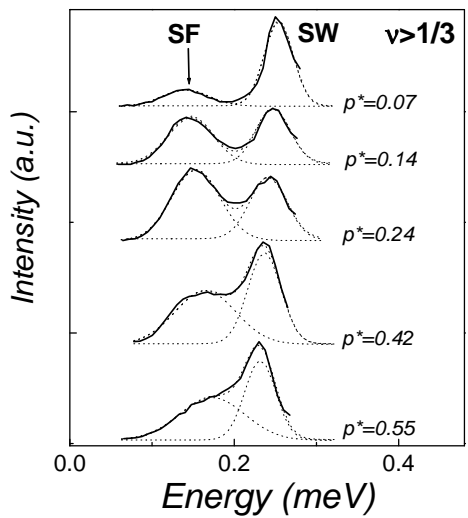,width=0.7\linewidth, clip=}
\caption{Evolution of the light scattering spectrum with $p^*$ for
$\nu>1/3$. The dotted line shows the analysis with of a
two-gaussian fit of the spin-flip (SF) and spin-wave (SW) peaks.}
\label{fig3a}
\end{figure}
\begin{figure}
\centering \epsfig{figure=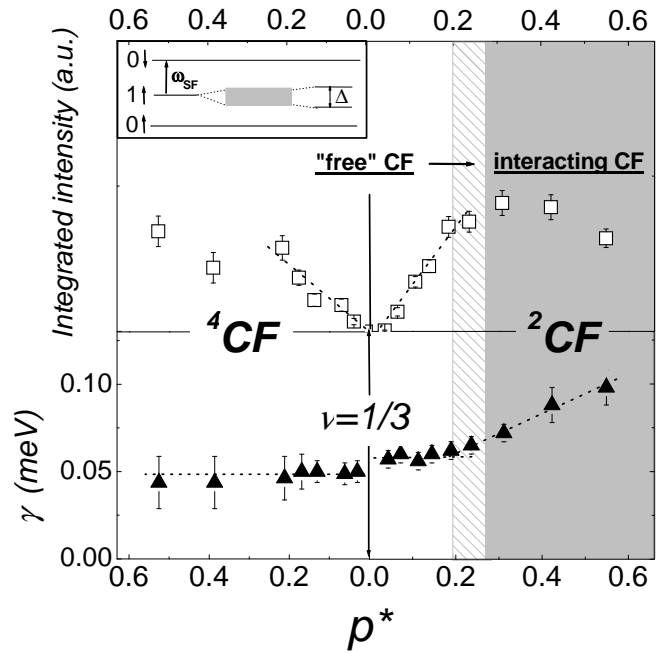,width=0.99\linewidth,clip=}
\caption{Evolution of the integrated light scattering intensity
(open squares) and the spectral linewidth $\gamma$ (full
triangles) of the spin-flip (SF) excitation as a function of
partial filling factor $p^*$. The left panel shows the data for
$\nu<1/3$ or $^4$CF, while the right panel shows data for
$\nu>1/3$ or $^2$CF. The dotted lines are guides to the eyes. A CF
Landau level scheme is drawn in inset. It depicts a simple picture
of the transition from "free" to interacting CF regime.}
\label{fig3}
\end{figure}

Figure \ref{fig3a} displays the evolution of the spectra  of SF and
SW modes for $\nu>1/3$ as a function of $p^*$. These spin excitations are
linked to ($^2$CF) quasiparticles. It is immediately apparent from
these spectra that the SF mode lineshape and energy are strong
functions of $p^*$. We constructed reasonably good fits to the
spectra with two gaussian functions, as shown in Fig. \ref{fig3a}. A
similar procedure was carried out for $^4$CF when $\nu<1/3$ (not
shown in Fig. \ref{fig3a}). The linewidth $\gamma$ and the
integrated intensity of the SF mode that are obtained by this
analysis are shown in Fig. \ref{fig3}.


The dependence of the integrated SF intensity and $\gamma$ on $p^*$
in Fig. \ref{fig3} reveals two distinct regimes. At low $p^*$, the
integrated intensity grows linearly with $p^*$ and the linewidth
remains constant around 0.06~meV$\pm$0.01. These features are
consistent with a picture of "free" CFs that applies to both $^2$CF
and $^4$CF sides. For $p^*>0.2$, however, the saturation of the
integrated intensity seems to indicate the onset of a breakdown of
the "free" CF picture.
\par
The observed $p^*$-dependence of the SF linewidths is consistent
with the scenario of an onset of a regime dominated by CF
interactions for $p^*>0.2$. The linewidth of the SF light
scattering peak of $^2$CF, shown in in Fig. \ref{fig3}, increases
when $p^*>0.2$ to up to 0.1~meV for $p^*=0.55$. The enhanced SF
linewidwth could be regarded as a key manifestation of the onset
of CF interactions. By contrast the SF peak remains sharp for $^4$CF,
likely due to the expected weaker residual interactions among $^4$CFs.
\par
The picture we propose to interpret these results assumes that
states of CF change with interactions in the same way as states of
electrons. In a free CF picture, such as the one in Fig.
\ref{fig1}, the partially populated $\ket{1,\uparrow}$ level is
expected to be infinitely sharp and CF quasiparticles are
long-lived. This picture is expected to hold at low $p^*$. As
$p^*$ increases CF quasiparticles acquire a finite lifetime due to
interactions and the $\ket{1,\uparrow}$ level is expected to
broaden significantly as sketched in the inset to Fig.
\ref{fig3}. In this regime the width of the SF mode is expected
to increase as seen in our data for $p^*>0.2$. The
$\ket{1,\uparrow}$ level may eventually split into well-defined
higher order CF Landau levels which, in turn, can give birth to
higher order FQHE states like the one observed at $\nu=4/11$ by
Pan et al. \cite{Pan}.
\par
An estimate of the CF level broadening can be obtained from the
evaluations of the gap $\Delta$ of the incompressible state at
$\nu=4/11$. Recent calculations, that incorporate corrections to
account for the width of the quantum well, give a value of roughly
0.1~meV \cite{Goerbig1}. This value is consistent with the
linewidths observed and supports the interpretation that the width
of the SF peak indeed increases due to the increasing impact of the
CF interactions for $p^*>0.2$.
\par
Additional insights can be obtained from the filling factor
dependence of the SF mode energy. Up to now, we have implicitly
considered that the level ordering is the one depicted in
Fig.\ref{fig1} and Fig.\ref{fig3}. This ordering assumes full spin
polarization in the full range $1/3<\nu<2/5$. While this assumption
has been debated \cite{Wojs-Quinn,Wojs-Quinn2,Mandal-Jain,Jain-chang},
 transport experiments with
tilted magnetic fields indicate that at least the state at
$\nu=4/11$ is fully spin polarized for $B_T>10~T$.
\par
In the following we show that the filling factor dependence of the
SF peak energy clarifies this issue. If we neglect excitonic
interactions between the excited CF quasiparticle and the
quasihole, one can write a simple expression for the SF mode
energy \cite{Mandal-Jain,Irene1}
\begin{equation}
\hbar\omega_{SF}=E_z^*-\hbar\omega_{CF}
\label{SF}
\end{equation}
where E$_z^*$=E$_z$+E$^{\uparrow\downarrow}$ and
E$^{\uparrow\downarrow}$ is the spin reversal many-body energy for
composite fermions \cite{Mandal-Jain-2}. E$^{\uparrow\downarrow}$
depends strongly on the spin polarization of the system and should
vanish in the absence of spin polarization. Therefore any loss of
spin polarization should drastically decrease the spin reversal
many-body energy and drive the SF peak energy towards zero energy
\cite{Irene3}.
\begin{figure}
\centering \epsfig{figure=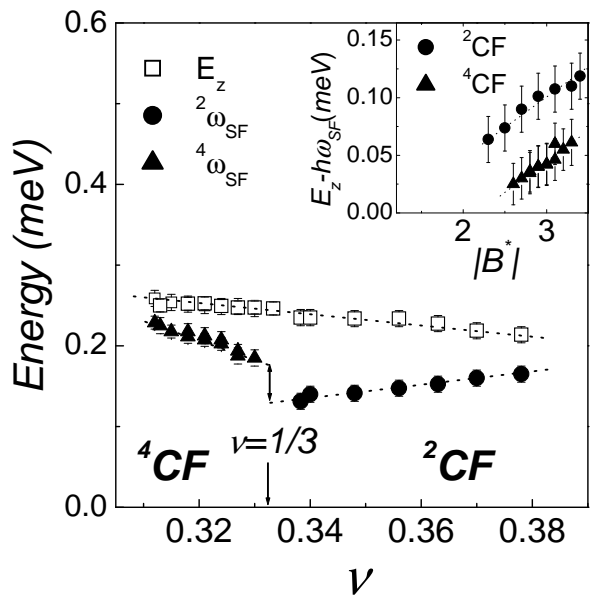,width=0.89\linewidth,clip=}
\caption{Filling factor dependence of $\omega_{SF}$ (Black full
circles and up-triangles for $^2\omega_{SF}$ and $^4\omega_{SF}$
respectively) and E$_{z}$ (open squares) for $0.31<\nu<0.38$. The
dotted lines are guides to the eyes. The inset shows the splitting
between E$_z$ and $\omega_{SF}$ plotted as a function of the reduced
effective field, $B^*$, of CF theory (see text).} \label{fig4}
\end{figure}
\par
The filling factor dependence of the SF mode energy is shown in
Fig.\ref{fig4} for $0.31<\nu<0.38$. The SF mode energies have
monotonic dependence on filling factor for both $\nu>1/3$ and
$\nu<1/3$. These behaviors clearly indicate that there is no
significant loss of spin polarization in this filling factor range.
\par
It is significant that the SF mode energies show opposite trends
with $\nu$ (or $B$): while $^2\omega_{SF}$ increases with $\nu$,
$^4\omega_{SF}$ decreases. The CF framework captures the opposite
trends of the SF mode energies with $\nu$. Indeed, according to
Eq. \ref{SF} and neglecting the weak $\sqrt{B}$ dependence of the
spin-reversal many-body energy, the splitting between the Zeeman
energy and $\omega_{SF}$ should scale as
$\left|{B^*}\right|$. $B^*$, the reduced effective field
experienced by CF quasiparticles, is written as $B$-$B_{1/2}$ and
$B$-$B_{1/4}$ for $^2$CF and $^4$CF respectively. The evolution of
the splitting as a function of $\left|{B^*}\right|$ is shown in
the inset of Fig.\ref{fig4}. These results are remarkably
consistent with key predictions of the CF framework in that the
mode energy is proportional to $\left|{B^*}\right|$ for both SF
excitations.
\par
A sharp discontinuity in the SF mode energy occurs at $\nu=1/3$.
This observation finds a natural explanation within the CF
framework because when crossing the $\nu=1/3$ boundary there is an
abrupt change in character of the quasiparticles in the ground
state of the quantum fluid. The discontinuity of the SF energy at
exactly $\nu=1/3$ is thus regarded as a manifestation of the
different energy scales associated with $^2$CF and $^4$CF
quasiparticles. Indeed, the energies of SF excitations are linked
to the CF cyclotron frequency $\omega_{CF}$ and the spin
reversal many-body energy E$^{\uparrow\downarrow}$. These energies
are expected to be markedly smaller for $^4$CF quasiparticles, as
found in the recent light scattering observations of $^4$CF
spin-conserving excitations \cite{Cyrus}. We note that a similar discontinuity has been 
reported recently in luminescence experiments around $\nu=1/3$ \cite{Byszewski}. 
\par
In summary, we have observed CF SF excitations away from $\nu=1/3$
and use them as a tool to study CF interactions away from major fractions.
The evolution of the lineshape of the SF peak excitation reveals a striking transition from a weakly
interacting to a strongly interacting CF regime. The breakdown of
the weakly interacting CF picture is expected to result in the
formation of higher order Fractional Quantum Hall states that do
not belong to the primary CF sequence. The results of light scattering experiments
in conjunction with CF theories that go beyond the mean-field
non-interacting limit should reveal intriguing new physics of
composite quasiparticles in the regime of the FQHE.
\par
This work is supported by the National
Science Foundation under Grant No. NMR-0352738, by the Department
of Energy under Grant No. DE-AIO2--04ER46133, and by a research
grant from the W.M. Keck Foundation.

\end{document}